\documentstyle[twoside,fleqn,espcrc2]{article}
\def\noi{\noindent}

\def \GG {{\bf \G}} 
\def \STr {{\rm STr }}
\def \Str {{\rm Str }}

\def \FF {{\rm F}} 
\def \SSTr{  {\widehat {\rm STr} } }
\newcommand{\rf}[1]{(\ref{#1})}
\def \diag {{\rm diag}}
\def \gym {g_{\rm YM}} 

\newcommand{\AmS}{{\protect\the\textfont2
  A\kern-.1667em\lower.5ex\hbox{M}\kern-.125emS}}

\def \hh {\lambda}

\def \four{{\textstyle {1\ov 4}}}
 
\def\bi {\bibitem}

\def \F {{\cal F}}

\def \del {\partial}

\def \ha{{\textstyle{1\over 2}}}

\def \a {\alpha}

\def \chi {\chi}
\def \s {\sigma}
\def \p {\phi}
\def \m {\mu}

\def \td {\tilde }

\def \ci {\cite}

\def \C {{\cal C}}

\def \inv {^{-1}}
\def \ov {\over }
\def \four{{\textstyle{1\over 4}}}

\def \V {{\cal V}}

\def \pa {\Vert}

 \def \G {{\Gamma}}
 \def \C {{\cal C}}

\def \foot{\footnote}

\def \Tr {{\rm Tr }}
\def\np {{  Nucl. Phys. }}
\def \pl {{  Phys. Lett. }}

\def \pr  {{ Phys. Rev. }}

\def \cqg {{ Class. Quant. Grav. }}
\def \bi{\bibitem}

\def \T {{\cal T}}
\def \tr {{\rm tr }}

\def\det{\hbox{det}}
\def\be{\begin{equation}}
\def\ee{\end{equation}}
\def\beq{\begin{equation}}
\def\eeq{\end{equation}}
\def\bea{\begin{eqnarray}}
\def\eea{\end{eqnarray}}

\def \la{\label}
\def \ci{\cite}

\def \FF {{\rm F}} \def \T {{\td T}}

\title{
{\small
\rightline{Imperial/TP/96-97/66}
\rightline{hep-th/9709123}
\rightline{September 1997}}
\vspace{1truecm}
Interactions Between  Branes and  Matrix Theories }
\author{A.A. Tseytlin\address{Theoretical Physics Group, Blackett Laboratory \\ 
        Imperial College, London SW7 2BZ, U.K. }
   \thanks{Contribution to the Proceedings of  Strings '97, 
Amsterdam, June 16-21, 1997.}
}
       
\begin{document}

\begin{abstract}
We review some  tests of the 0-brane and instanton  matrix models 
based on comparing  long-distance  interaction potentials 
between branes  and their bound states (with 1/2,1/4 or 1/8 of supersymmetry)
in supergravity and in super Yang-Mills descriptions. We first
consider the supergravity-SYM correspondence at the level
of the leading term in the  interaction potential, and then describe some  recent results concerning the subleading term and  their implications for the structure of the 2-loop $F^6$ term in the  SYM effective action.
\end{abstract}

\maketitle

\section{Introduction}
Below we shall review  some  recent results  about  the 
correspondence  between type II supergravity and matrix theory
(or super Yang-Mills) 
descriptions of long-distance interactions of 
certain  p-branes \ci{CT1,CT2,bbpt,CT3} (see also 
\ci{be,mal}).
 We shall emphasize   the common features   underlying 
the agreement  between the two pictures 
for  different 
 brane configurations with varied  amounts of supersymmetry.
 
One of the  ideas  behind  the  Matrix theory proposal \ci{bfss} 
(considered in weak-coupling limit) is that 
one should treat 0-branes as fundamental, effectively building
other branes out of large numbers of 0-branes. 
That this  is possible in principle
follows from the existence of open string theory description 
of D-branes, i.e. 
from  T-duality relating a system of D0-branes on a torus $T^p$
to Dp-branes wrapped over the dual torus $\T^p$ \ci{poll,tay}.
Similarly, one may consider  D-instantons as
 basic  building blocks  for  D-branes of type IIB theory. 
The 
clusters  of $N$ D0-branes or $N$ 
 D-instantons 
 may be described  (at low energies) 
  by $U(N)$  SYM theories obtained by reduction from   $10$
 to  $1$ or $0$ dimensions \ci{wit2}, 
which then define   the  corresponding 
`0-brane'  \ci{bfss}  and `instanton' \ci{ikkt}
matrix models. 

Starting with the instanton model ($S_{-1} = {1 \ov 4 g_s} \tr [X_a, X_b]^2 +
$fermionic terms)  and expanding near the vacuum
$\bar X_0 = \diag(\td x_0^{(1)}, ...,\td x_0^{(N)})$, \ $\bar X_m=0$, 
 corresponding to the 
instantons distributed along the euclidean time direction, one may 
relate $S_{-1}$ to the 0-brane matrix model action
 by using that in the large $N$ limit $\bar X_0 \to  i \del/\del x_0$ \ci{ikkt}.
An example of a  non-trivial classical solution ($[X_a, [X_a, X_b]]=0$)
is provided by $[\bar X_a, \bar X_b] = i \FF_{ab} I_{N\times N}$, 
or, e.g.,  $\bar X_a = i \del_a +  A_a, \  A_a = - \ha \FF_{ab} x_b, $
where  $\FF_{ab}$ ($a,b= 0,..., p$) are non-vanishing 
constants.  Such backgrounds describe  non-marginal bound states
$p+(p-2) + ... + (-1)$ of type IIB  Dp-branes with other branes.
Similar configurations  ($F_{ab}\to F_{mn}$, \ $m,n= 1,...,p$)
in the 0-brane matrix model  \ci{bfss,grt,bss} 
describe  type IIA  1/2 supersymmetric 
non-marginal bound states of branes
$p+(p-2) + ... + 0$.
For example,  a  `D-string' in  the instanton matrix model ($p=1$),  
i.e.  a  D-string bound to D-instantons \ci{nbi}, 
 is  T-dual to $2+0$ bound state in type IIA theory, or `longitudinal M2-brane'.
BPS states with 1/4 of supersymmetry ($3\pa (-1)$ or $4\pa 0$) 
  may be represented by $\bar X_a$ with self-dual commutator 
$[\bar X_a, \bar X_b]$ \ci{doug,grt,bss}.

Here we shall  consider only  branes wrapped over tori, i.e. 
the case of compactified matrix theory which is described by 
the  SYM theory on the dual torus  \ci{bfss,tay,grt,sussk} ($\T^{p+1}$ in IIB case 
or $\T^{p}$ in IIA case). Thus instead of representing, e.g., IIA 
brane backgrounds
in terms of large $N$ matrices (or differential operators) 
of $D=1$  SYM theory (as was done in \ci{bfss,ab,lifmat,lif3,CT1,CT2})
we shall use equivalent but more  straightforward  representation 
in terms of $D=p+1$ SYM backgrounds.
The non-marginal 1/2 supersymmetric 
bound states of branes 
are then described by  constant magnetic backgrounds $F_{mn}$, 
while marginal and non-marginal 1/4 supersymmetric 
bound states  (e.g., $1\pa 0$ and $4\pa 0$)
-- by  1/2 supersymmetric BPS states of SYM theory
(wave $F_{m+}=0$ and instanton $F_{mn}=F^*_{mn}$) 
 and their superpositions with magnetic backgrounds, see, e.g, \ci{guraln,dvv,gopak,taylor}.
The 1/8 supersymmetric bound states of 
branes may be described by more general  1/4 supersymmetric 
SYM configurations  which are superpositions
of the  wave and/or instanton backgrounds
 (see \ci{dps,limart,halyo,dvvv,malda,CT3}
and below).

In certain cases of BPS bound states of branes
(having  non-trivial 0-brane content
in IIA case or instanton content in IIB case)\foot{Equivalently, 
the required configuration of branes should be reducible (by T-duality)
to  two parallel (wrapped over $T^p$) 
  bound states  of p-branes with other branes, 
e.g., $4+2$ and $4\pa 0$, which  should have SYM description.}
 one  may expect a 
 correspondence between their 
description   in terms of curved supersymmetric backgrounds
 of  type II supergravity compactified on $T^p$ 
   and their 
description as SYM backgrounds on the dual torus $\T^p$.
On SYM side, though it may seem that one may  not actually need to 
assume 
that $N$ is large in order to have this correspondence \ci{suss}, 
 one, in fact,  is to consider only planar diagrams
corresponding to large $N$ limit. 
  As discussed in \ci{bbpt,CT3}, on supergravity side,  
this corresponds to viewing IIA configurations of branes as resulting
from a $D=11$ theory in which a null direction 
$x_-= x_{11}-t$ is compactified \ci{suss}. This is formally equivalent
to the prescription of computing the interaction potentials between branes
by taking the 0-brane harmonic function 
without its standard asymptotic value 1 (see below).
 As we  shall see, 
a similar $H \to H-1$  prescription applies in the  IIB instanton 
case.

On the SYM side,  the interaction potential between two different  BPS 
configurations  of branes is represented by the SYM effective action $\GG$
 computed 
in an appropriate SYM  background \ci{dkps,bfss,ikkt,ab,lifmat}. 
More precisely, the effective action in question  should be  obtained 
by integrating out  only massive  excitations in a  background representing separated clusters of branes. However, in the case  when each cluster
is a BPS state with  vanishing 
self-energy corrections one may not distinguish 
the full effective action from the "interaction" part of it. 
 In general,  both  the vectors $A_a$ ($a=0, ..., D-1$) and
the scalars  $X_i$ ($i= D, ...,9$) 
may have  non-trivial  background values. 
Consider, for  example, 
 a system  of 
a $0$-brane  probe 
interacting with a BPS bound state of branes (wrapped over $T^p$), 
containing, in particular,  $N_0$\  0-branes.
   Under  T-duality this  becomes a system of a 
  Dp-brane probe with charge $n_0$ and  a Dp-brane  source with charge $N_0$ bound to some
 other branes of lower dimensions (both probe and source being 
wrapped over $\td T^p$).
If the probe and the  source are separated by a distance $r$ 
in  the direction 8 and the probe has velocity $v$ along the transverse 
 direction 9, 
this configuration  may 
   be described by the following  $u(N)$, $N=n_0 + N_0$, SYM background
on $\td T^p$:  
$
\bar {A}_a=
 \pmatrix{ 0_{n_0\times n_0}  & 0 \cr 0 & {A}_a \cr}$ ,  $\ \ \bar {X}_i=
 \pmatrix{
0_{n_0\times n_0}  &0 \cr 0 & {X}_i \cr } ,$ \ 
$\bar X_8=
 \pmatrix{ r\ I_{n_0\times n_0}   &  0 \cr 0  &  0_{N_0\times N_0} \cr}
,$ \ $ 
 \bar X_9= \pmatrix{  vt\ I_{n_0\times n_0}   & 0  \cr 
   0 & 0_{N_0\times N_0} \cr }  , $
 where ${A}_a$ and $X_i$ are $N_0 \times N_0$ matrices in the 
fundamental representation of $  u(N_0)$
 which  describe the source bound state.
The  dependence on  derivatives of the scalar fields 
$X_i$ may be formally determined from the dependence of
the effective action 
 $\GG$ on the gauge field
 in a  higher-dimensional background  representing 
T-dual ($X_i \to A_i$)  configuration.
In particular,  the dependence on  the   velocity $v$  
may be described  by a gauge field
  background $F_{09} \sim v$ \ci{bachas,ber,CT2}.
This is formally the same as an extra constant field strength background 
switched on the probe in  the euclidean D-instanton model 
 case.


The background value of $X_8$  plays the role of an IR cutoff $M\sim r$
in (the interaction part of) $\GG$ 
(in the open string theory picture it is related 
to  the mass of  the  open string states stretched between the probe and source branes).
  The long-distance interaction potential $\V$  will be 
 given  by the  leading IR terms  in the 
expansion of $\GG$ in powers of  $1/M$, or, equivalently, 
by the low-energy expansion in powers of  the gauge field strength 
$F$.

On the  supergravity side, the interaction potential 
may be determined   from  the  action
of a brane probe 
moving in  a  curved background produced by a 
brane source.
For example, 
 the  action  for a  0-brane probe  in  
 a background produced by a marginal bound state of branes $1\pa 0, \ 4\pa 0$, \ 
$4\bot 1 \pa 0$ or $4\bot 4\bot 4 \pa 0$  (which is essentially the same as  the  action 
for a $D=11$ graviton scattering off the corresponding  M-brane configuration
$2+$wave, $5+$wave, $2\bot 5+$wave or $5\bot5\bot5+$wave)
has the following general structure
\ci{ts3,ts4,dps}
$$
I_0 = -T_0 \int dt\ H_0\inv \big[\sqrt { 1 - H_0 H_{(1)} ...H_{(k)}  v^2} -1 \big]
$$
\be 
\equiv \int dt \ \big[ \ha T_0 v^2 - \V (v,r) \big] \ .
\la{pou}
\ee
Here $H_0$ and  $H_{(1)}, ..., H_{(k )}$ ($k=1$ for 1/4 supersymmetric bound states and
$k=2$ or $3$ for  1/8 supersymmetric bound states)  
   are the  harmonic functions  $H_{(i)}  = 1 + Q_i/r^{7-p}$
representing  the constituents of  the bound state.
Since  for D-branes \ci{poll}  $Q_i \sim g_s N_i$ and $T_0 \sim  n_0 g_s\inv$ 
($g_s$ is the string coupling constant),  the 
long-distance expansion of the classical 
supergravity interaction potential $\V$  has the following form\foot{For simplicity, here we are assuming 
 that the  bound state has only RR charges;
cases with non-vanishing  fundamental string charge or momentum
  ($Q_1 \sim g^2_s$) 
 can be treated in a similar way, see  \ci{dps,GR,CT3} and below.}
\be
 \V  = \sum_{L=1}^\infty \V^{(L)} = 
{ n_0 \ov g_s}  \sum_{L=1}^\infty 
 \big({ g_s  \ov r^{7-p}}\big)^{L}  {k_L} (v, N_i)  
\ ,  \la{vvv}
\ee
so that  the $(1/r^{7-p})^L$ term 
has the same $g_s$ dependence as in the $L$-loop
term in SYM theory with coupling $\gym^2 \sim g_s$.
The detailed  structure  of the coefficients  in \rf{vvv}  reflects
 the  special role played by the 0-brane function $H_0$ and 
the presence  of the product of the `constituent' harmonic functions
(which is a direct consequence of 
 the `harmonic function rule' structure \ci{ts2}
of the  supergravity backgrounds representing marginal 
BPS bound states of branes).

To have a precise agreement  between  the supergravity  and SYM
expressions for the 
potential (already at the leading level) 
one needs to assume that 
in expanding  \rf{pou} in powers of $1/r^{7-p}$ one should
set 
$
H_0 =  Q_0/r^{7-p}, \ Q_0 \sim g_s N_0$. 
This prescription 
may be interpreted in two 
possible ways (which are equivalent for the present purpose of comparing interaction potentials). 
One may  assume (as  was done in \ci{lifmat,lif3,CT1}) 
that $N_0$ is large  for fixed $r$ 
(larger than $N_1,..., N_k$ and any other charge parameters that may be present in non-marginal brane
configurations),
 so that $H_0= 1 + Q_0/r^{7-p} \approx Q_0/r^{7-p}$. 
Alternatively, one may keep $N_0$ finite but consider  the $D=10$ 
brane system  as  resulting  from an M-theory configuration 
 with 
 $x_- = x_{11} -t$ being compact \ci{suss}:  as  was 
pointed out  in \ci{bbpt}, the 
dimensional reduction of a  $D=11$ gravitational
wave combined with  M-brane configurations
along $x_-$  results in supergravity backgrounds with 
$H_0= Q_0/r^{7-p}$.  

The formal technical   reason why the 
leading-order SYM and supergravity 
potentials  happen to agree in certain simple  cases 
 is  related to the fact that the combination 
of $F^4$ terms that appears in the 1-loop
SYM effective action is also the  same as the one  in the  expansion
of the Born-Infeld action\foot{This fact is due to maximal supersymmetry of SYM 
theory: the 1-loop $F^4$ terms in the bosonic YM theory are different from the combination  in BI action.} \ci{MT}, but the 
 latter is  closely related to 
 the action of a D-brane probe moving in a supergravity background. 
This becomes especially clear in the type 
IIB (instanton model)  context  \ci{CT2}, provided one takes into account 
  that because of the T-duality involved, 
the relevant 
gauge field backgrounds which appear in the SYM and supergravity 
descriptions are related by  $F_{ab} \to (F_{ab})\inv$  \ci{CT2}.
(equivalently, one may consider directly the T-dual system
of p-brane parallel to  $p+...+ (-1) $
brane; in that case the gauge field on the brane and the SYM background
are related directly).

The  known (weak-coupling string theory) 
explanation \ci{dkps,lifmat}  of  
 the  precise agreement  between the 
leading-order   supergravity 
and 1-loop SYM potentials in certain simple cases 
   uses the observation that  for 
configurations of branes with  sufficient  amount of
underlying supersymmetry,  the 
long-distance and short-distance limits of the  string-theory 
potential  (represented  by the annulus diagram \ci{poll})  
are the same. That   implies that  the 
 leading-order 
(long-distance) interaction potential  determined by 
the    classical 
supergravity limit of the closed string theory 
is  the same as the  (short-distance) one-loop potential 
produced by  the massless (SYM) 
open string theory modes.

The  results of \ci{BB,bbpt,CT3}  suggest that this  supergravity-SYM 
correspondence  should extend beyond  the leading-order level.
One may conjecture that,  in general, 
the existence of the open string theory description 
of D-branes  combined with enough supersymmetry 
implies again the agreement between long-distance and short-distance
limits of higher open string loop  terms in the 
 interaction potential. 
Equivalently, that would mean that\  (i)  the leading IR part of 
 the $L$-loop  term in the  $SU(N)$ 
SYM effective action in $D=1+p$ dimensions  
has  a universal   $F^{2L+2}/M^{(7-p)L}$ structure, and \ \  
(ii) computed for a SYM background  representing 
a configuration of interacting branes, 
 the  $F^{2L+2}/M^{(7-p)L}$
  term  should reproduce the  $1/r^{(7-p)L}$ term 
in the corresponding classical supergravity  potential.

Below we shall first demonstrate  the  agreement between the 
leading-order 
terms in the SYM  and supergravity expressions for the interaction 
 potential (section 2) and then discuss 
what is known about that correspondence at the 
level of subleading  terms  \ci{bbpt,CT3} (section 3).
Some concluding remarks will be made in section 4.

\section{ Leading-order interaction   potentials}
\subsection{ SYM effective action}
In general, the  effective action 
of the  $D=p+1$ dimensional  $U(N)$ SYM  theory on $\T^p$ 
($ S = - { 1 \ov 2 \gym^2}  \int d^{p+1 } \td x\  \tr F^2 + ...,$ \ $
 \gym^2 = (2\pi)^{-1/2} g_s \td V_p$)
 for   a purely  gauge field 
background
 and  with an explicit 
 IR cutoff $M$ has the following structure\foot{We will  consider  only in the 
low-energy  limit of the SYM  theory, i.e. will not  consider the 
 UV  cutoff dependent parts in the corresponding effective 
actions (assuming  the existence of 
an explicit cutoff effectively provided  at  weak coupling  
by the string theory).}
$$
\GG =   \sum_{L=1}^\infty  (\gym^2 N)^{L-1} 
\int  d^{p+1} \td x  \sum_n  {  c_{nL} F^{n}  \ov  M^{2n - (p-3) L -  4}}  
\ .  $$
 We  will  be interested   only  a  special   subset 
of terms in  $\GG$ 
(generalising the `diagonal terms' in \ci{bbpt}) 
which have the same  coupling $g_s$, 0-brane charge $N_0$ 
 and distance $r=M$ dependence as 
the terms in the long-distance expansion \rf{vvv}  of the classical 
supergravity interaction potential $\V$ between  a Dp-brane probe
(with tension $\sim n_0/g_s$) and a Dp-brane source (with charge parameter
$\sim g_s N_0$).
We shall assuming that the SYM backgrounds describing individual branes
are supersymmetric, so that the effective action vanishes  when evaluated on
 each of them  separately (i.e. its non-vanishing part will represent the interaction between branes).

One may  conjecture   that 
due to  maximal underlying  supersymmetry of the SYM  theory, 
the terms $F^{2L+2}/M^{(7-p)L}$ represent, in fact,  the {\it leading} 
IR (small $F$ or, equivalently,  large $M$) contribution
 to $\GG$ at   $L$-th loop order.
 This is indeed true for $L=1$ \ci{MT}
and, in view of the results of 
\ci{BB} (for $p=0$) and \ci{seid} (for $p=3$) this 
  should    be  true  also for $L=2$.
The sum of  such  leading IR   terms   at each loop order 
 will be denoted as  $\G$. Thus 
$$
\G  = {1 \ov 2 \gym^2 N}  \sum_{L=1}^\infty  \int d^{p+1} \td x 
  \  \big( {a_p  \gym^2 N \ov  M^{7-p}}\big)^L  
    \hat  C_{2L+2} (F)  , 
$$
where $\hat  C_{2L+2} (F) \sim  { F^{2L+2} }$  and $a_p$ are universal coefficients not depending on $N$ or $L$.

At the 1-loop level, $\GG^{(1)}= \G^{(1)} +  O\big(\frac{1}{M^{9-p}}\big)$, 
where
\be
\G^{(1)}
=   \frac{ a_p } {   2 M^{7-p} }\int d^{p+1} \td x \ \hat  C_4 (F)
\ , \    \la{hhh}
\ee
$$
\hat  C_4 = \STr\  C_4 
= - {\textstyle{ 1 \ov 8}} \STr \big[ F^4 -\four (F^2)^2 \big] 
$$
$$
 = \ -  {{\textstyle{1\over 12}}} \Tr \big(F_{ab} F_{bc} F_{cd} F_{da}
+ \ha  F_{ab} F_{bc}   F_{da} F_{cd}   
$$ \be - \  \four
F_{ab} F_{ab} F_{cd} F_{cd}  - {{\textstyle{1\over 8}}}
F_{ab} F_{cd}  F_{ab} F_{cd} \big) \,. 
\la{bbb}
\ee
Here $
  a_p = 2^{2-p}\pi^{-(p+1)/2}
{\Gamma ({\textstyle{ 7-p \ov 2}})} $  and 
$\STr$ is the symmetrised trace in the adjoint representation
 (for $SU(N)$ \  $  \Tr Y^4 = 2N \tr Y^4 + 6 \tr Y^2 \tr Y^2$,  
    so one gets the expression  containing  terms with single and double traces in the fundamental representation  \ci{CT2}).
The   polynomial $C_4$  is the same
one  that appears in the expansion of the BI action, 
\be
\sqrt{-\det{(\eta_{ab} +   F_{ab} })} 
= \sum_{n=0}^\infty     C_{2n} (F)   \ ,  
\la{see}
\ee
$$
C_0=1\ , \ \  \ \ \ C_2 = -  { 1\ov 4} F^2  \ , \ \ \ 
C_4 = - { 1\ov 8}
 \big[F^4 - { 1\ov 4} (F^2)^2 \big]  \ , 
$$
$$
C_6 = - { 1\ov 12} \big[F^6 - { 3 \ov 8 } F^4 F^2  
+ { 1 \ov 32} ( F^2)^3\big] \ ,  \ \ ....  \ ,  
$$
where 
$F^2= F_{ab}F_{ba}\ , \ \ $ $ F^k = F_{a_1a_2}
 F_{a_2a_3} ... F_{a_ka_1} . $

\subsection{ Potentials from SYM theory }
Let us now consider several examples of different  brane configurations
  which  admit  a SYM description and 
compute the leading-order 
potentials $\V^{(1)}$  by substituting 
the   corresponding gauge field backgrounds into \rf{hhh}. 
We shall assume that $ \G^{(1)} = \int dt \V^{(1)}$ in the cases involving 0-branes  and   that 
$ \G^{(1)} =
 \V^{(1)}$ in the  D-instanton `interaction' cases. 
In what follows  we shall set  $2\pi \a'=1$  and assume for simplicity
 that 
the volumes of the tori take self-dual values, 
$V_p = \td V_p = (2\pi)^{p/2}$.

In the 1/2 supersymmetric case the basic  example is 
the interaction of  two  parallel  non-marginal   IIB 
bound states  $p + (p-2) + ...+ (-1)$ separated by a distance.\foot{One may  
consider also  interactions between 
orthogonally oriented branes,  
assuming that they are wrapped around
parts of common torus.}
  They 
 may be  represented by  the following 
$U(N)$, $N= n_{-1} + N_{-1}$,  background on the dual torus $\T^{p+1}$
($n_{-1}$ and $N_{-1}$ are the instanton numbers of the two branes
on $T^p$ or the numbers of Dp-branes on $\T^p$):

\noindent
$   X_9 = \diag ( r I_{n_{-1} \times n_{-1}}, \ 0_{N_{-1} \times N_{-1}} )$, 

\noindent
$   F_{ab} = 
 \diag (\FF_{1ab} I_{n_{-1} \times n_{-1}}, \ \FF_{2ab} I_{N_{-1} \times N_{-1}} )$, 

\noi
where $F_{1,2}$ are constant  parameters describing the charges
of the two  bound states.  
The  $su(N)$ analogue of $   F_{ab}$
(its traceless part) may be written as 
$$   F_{ab} = \FF_{ab} J_0 \ ,  \ \ \ \FF\equiv  \FF_1 - \FF_2  \  , 
\ \ $$ $$ 
J_0 =    {\textstyle{1\ov n_{-1} + N_{-1}}}\diag (N_{-1}
   I_{n_{-1} \times n_{-1}}, \  -n_{-1}
  I_{N_{-1} \times N_{-1}} ) \ . $$
 Since this background is  abelian, 
$\STr$ is equivalent to $\Tr$ and thus the coefficient $\hat C_4$ in \rf{hhh}
determining the interaction  potential is  simply 
\be
\hat  C_4 
= - {\textstyle{ 1 \ov 4}}  n_{-1} N_{-1} \big[ \FF^4 -\four (\FF^2)^2 \big]
  \  . \la{ess}\ee
A particular example   is that of the interaction 
 between an  instanton  and  a  $p + (p-2) + ...+ (-1)$
state (here $\FF_1=0$) \ci{CC,CT2}.

Similar result is found  in  the type IIA case,  for example,  for 
a 0-brane (with velocity $v$) 
interacting 
with $p + ... +0$  IIA bound state described by

\noi 
$   F_{mn} = 
 \diag ( 0_{n_{0} \times n_{0}}, \ \FF_{mn} I_{N_{0} \times N_{0}} )\ 
(m,n=1,..., p$).
The IIB and IIA two cases are formally related by 
$ \FF_{ab} \to  ( \FF_{09} = v, \ \FF_{mn})$, so that here  \ci{CT1,CT2}  
\be
\hat  C_4 
= - {\textstyle{ 1 \ov 4}} 
 n_{0} N_{0} \big[ \FF^4 -\four (\FF^2)^2 
    -   v^2 \FF^2   + v^4 \big]   . 
\la{yyy}
\ee
Special cases, e.g.,  0-brane - 0-brane  ($\FF=0$)  and 0-brane - (2+0)-brane  ($\FF_{12} ={n_2\ov n_0}$)  interactions  were  discussed in \ci{ab,ber,lifmat,lif3,ballar}.  

An example of a  bound state with 1/4 of  supersymmetry 
is $4\pa 0$ which may be described \ci{doug} 
by a   self-dual  $SU(N)$ background 
on $\td T^4$:

\noi
 $F_{mn} =  F^*_{mn}, \  \int d^4 \td x \tr (F_{mn} F_{mn}) = (4\pi)^2 N_4$,

\noi
 or, explicitly, by a constant   background:

\noi
$F_{12} = F_{34} = q J_1 ,  \ \ q^2= {N_4\ov N_0} , $

\noi      $ J_1 \equiv 
\diag(0_{n_{0} \times n_{0}}, 
\ I_{\ha N_{0} \times \ha N_{0}},  \ - I_{\ha N_{0} \times \ha N_{0}} ) 
 $. 

\noi
Since the resulting background is commuting, the potential is again
given essentially by \rf{ess} or \rf{yyy}.
For example,  in the case of the $(4\pa0)-(4\pa0)$ 
system of two parallel $4\pa0$ states
with charges $(n_4,n_0)$ and $(N_4,N_0)$ 
we find \ci{CT1}
\be
\V^{(1)} = -{ n_0 N_0  \ov 16 r^3} [ 4 ({ N_4\ov N_0}  + {n_4\ov n_0} ) v^2 
 +  v^4 ]   .
\la{ff}
\ee
Similarly, for the  static potential between orthogonally oriented (within 6-torus)
$2+0$  and $4\pa 0$  states we get \ci{CT1}
$$
\V^{(1)} = -{ n_0 N_0  \ov 16r} [  ({n_2\ov n_0})^4
  - 4  ({n_2\ov n_0})^2 {N_4\ov N_0} +  O( v^2)  ]   .
$$
Analogous  expressions are found in the case 
of $1\pa 0$ bound state of a fundamental string and a 0-brane
which is describes by a plane wave background 
$X_2= X_2 (\td x_1 + t)$ or, in the T-dual picture, 

\noi
$A_2= A_2 (\td x_1 + t)$, \ $ F_{12}=F_{02} = h I_{N_0 \times N_0}$,

\noi
 where $h=h(\td x_1 +t)$ is a periodic function 
normalised so that 
$< h^2 > = g_s {  N_1\ov N_0} $, where $N_1$ is the string winding number
(see \ci{GR,CT3}). for example, for a 0-brane interacting 
with $4\pa 0$ we get the expression similar to the $0-(4\pa 0)$ case
(i.e. \rf{ff} with $n_4=0$)
$$
\V^{(1)}  = - { n_0 N_0 \ov 8 r^6} ( 4 <h^2> v^2 + v^4) \ . $$
Similar  expressions  describe also
interactions involving  the corresponding  T-dual 
type IIB bound states $3\pa (-1)$ and $(-1)+$wave.

To  determine  the  leading-order 
potentials  for configurations involving 1/8 supersymmetric 
states one  needs  to find their SYM description
and substitute the resulting backgrounds into \rf{hhh}.
The configuration of a 0-brane interacting with $4\bot 1 \pa0$  state
wrapped over $T^5$ (corresponding to extremal $D=5$ black hole)
 may be described by  a combination  of an instanton and a momentum wave
(carried, in general, by  vectors {\it and} scalars), 
or, explicitly
(after T-duality trading scalar backgrounds for the vector ones) \ci{CT3}\foot{This `instanton+wave'  $u(N)$ gauge field  background,  which should be
  representing the
marginal BPS  $5\pa 1 +$wave
configuration invariant
under 1/8 of ${\cal N}=2,D=10$  type IIB supersymmetry,
is,  indeed,    preserving  1/4 of the ${\cal N}=1,D=10$ 
 supersymmetry  of the SYM theory \ci{CT3}.}
 \be 
   F_{09} = v J_0 \ , \  \  \ \ \ \ \  \ 
   F_{12}=   F_{34} = q J_1 \ ,   \la{trat} \ee
\be
   F_{51} =    F_{01} =h J_0  \ , \ \ \ \ \ 
   F_{56} =    F_{06} =w J_0   \   ,  
\la{tat}
\ee
where the $su(n_0 + N_0)$ matrices   $J_0, J_1$  were  defined  above, 
$q^2= {N_4\ov N_0}$, 
and  the periodic  `vector wave' and `scalar wave' 
 functions $h=  h(\td x_5 +t)$ and $w= w (\td x_5 +t) $  
satisfy

\noi
$  < h^2 + w^2 > = { 1 \ov \td L_5} \int d \td x_5\  ( h^2  + w^2) 
= g_s {N_1\ov N_0}, $ 

\noi
i.e. $ < h^2 + w^2 > $ is proportional to the total  momentum 
of the wave in the  T-dual ($5\pa 1 +$wave) configuration.
The $\hat C_4$-coefficient of the corresponding  leading-order 
potential is found to be (equivalent 
 results were obtained in \ci{dps,malda})
\be 
\hat C_4 = - { 1 \ov 4}  n_0 N_0 \big[ 4 v^2 q^2  + 4 v^2( h^2 +w^2)  + v^4\big] \ .
\la{one}
\ee
The same  expression is found   for  the 0-brane interaction
with $4\bot4\bot4\pa 0$ bound state wrapped over $T^6$ (corresponding
to extremal $D=4$ black hole), or for  $6-(6\pa 2\bot2\bot2)$  interaction
on $\T^6$.  The $6\pa 2\bot2\bot2$  configuration 
may be described  by an `overlap' of the three 4d instantons on $\T^6$, i.e. 
by  the following   $su(N_0)$ constant gauge field strength 
 background \ci{CT3}
 \be
F_{14} =  F_{23} = q_1 \hh_1  \  , \ \ \ 
F_{45} =  F_{36} = q_2 \hh_2 ,     \la{weew}
 \ee
 $$
F_{15} =  - F_{26} = q_3 \hh_3  \ ,  $$
 where  $
 q^2_k =   {N_{4(k)}  \ov N_{0}}$
 and  $\hh_k$ are some 
 three   independent $su(N_0)$  matrices
normalised so that this gauge configuration
 produces the right 2-brane charges (and only them)  on 6-brane.
One possible choice of $\hh_k$ is the following  {\it `commuting'} one
(assuming that $N_0$ is a multiple of 4):
$
\hh_k = \m_k \otimes I_{{N_0\ov 4} \times {N_0\ov 4}}$, 
where $\m_k$ are the  diagonal $4\times 4$   matrices   (used  in \ci{taylor})
$\m_1 = \diag( 1,1,-1,-1) , $ $ 
\  \m_2 = \diag( 1,-1,-1,1) ,$  \  $ \m_3 = \diag(1, -1, 1,-1) $.
A {\it `non-commuting'} choice is to set $\hh_k$ to be proportional to the Pauli matrices $\s_k$, i.e. 
$
\hh_k = \s_k \otimes I_{{N_0\ov 2} \times {N_0\ov 2}}$.
 Both choices  represent  1/4 supersymmetric configurations 
 in the $D=6+1$ SYM theory \ci{CT3}. The leading-order potential 
in this case is proportional to
\be
\hat C_4 = - { 1 \ov 4}  n_0 N_0 \big[ 4 v^2 (q^2_1 + q^2_2 + q^2_3)  
  + v^4\big] \ .
\la{two}
\ee
Comparing  \rf{one},\rf{two} with  various special cases
discussed above  we conclude
 that, as might be expected, 
 the {\it leading-order}  SYM 
interaction potentials  for  {\it marginal } bound states
are essentially   the sums of pair-wise  interactions 
between constituent branes.
The same  will, of course,  be  true on the 
supergravity side (cf. \rf{pou}), and the potentials
 will  be in full agreement. 

It should be mentioned also that the leading-order interactions
involving non-supersymmetric bound states of branes \ci{taylor,pie,kraus}
or near-extremal
black holes \ci{malda}  are 
again described by the universal $F^4$  action \rf{hhh}.

\subsection{Potentials from supergravity}
To find the supergravity potentials we shall use the probe method, i.e. consider 
the action of a D-brane probe ($I_p = - T_p\big[
\int d^{p+1} x  e^{-\phi}  $ $ \sqrt {\det\big( G_{ab} +
 G_{ij} \del_a X^i \del_b X^j +  \F_{ab} \big)}-  \sum C  e^{\cal F} \big])$
in a curved background produced by 
a brane bound state as a source. 
 The key example in the 1/2 supersymmetric brane case is 
the interaction of a D-instanton  with  a type IIB 
non-marginal bound state $p+(p-2)+...+ (-1) $.
 The action for  the latter considered as a 
probe   may be  found  
by switching on a constant $\F_{ab}$  background
on the Dp-brane world volume \ci{ggut}.
The fluxes produced by $\F_{ab}$ determine the numbers of branes \ci{doug}
of each type in the bound state. In particular, the instanton number is

\noi
$
n_{-1} = {n_p  V_{p+1} \ov  (2\pi)^{(p+1)/2}} \sqrt {\det  \F_{ab}} = 
n_p \sqrt {\det  \F_{ab}}$

\noi
 (we assume that $T_p= n_p g_s\inv (2\pi)^{(1-p)/2}$
and $V_{p+1} = (2\pi)^{(p+1)/2}$). 
Substituting the D-instanton background
\ci{ggp} ($  ds^2_{10} = H^{1/2}(dx_a dx_a +  dx_idx_i )$, etc.,)
`smeared' along the directions of $T^{p+1}$
into the Dp-brane action and  ignoring the dependence of
$X_i$ on the  world-volume coordinates $x_a$ we find
$$ I_p  = - T_{p} V_{p+1} 
 H^{-1} $$
$$ \times \big[  \sqrt { \det ( H^{1/2} \delta_{ab} 
 + \F_{ab}) } - \sqrt {  \det \  \F_{ab}} \big] $$ $$ 
 =  - T_{-1}  \  H^{-1}  \big[ \sqrt { \det ( \delta_{ab} +  H^{1/2} \F^{-1}_{ab}) } - 1\big]   ,  
$$
where $ T_{-1} =   2\pi g_s\inv n_{-1}
  = T_{p} V_{p+1} \sqrt {  \det \  \F_{ab}}$.
Thus we got the BI action    with the field  
$\F_{ab} \to \F^{-1}_{ab}$ in a curved background. 

To establish the 
agreement with  the   leading-order potential in  the SYM  theory
one may  assume 
 that $\F_{ab}$ (and thus $n_{-1}$) is very large and expand
in powers of $\F\inv$ \ci{CT1}, or, alternatively,  drop 1 in the source 
D-instanton harmonic function $H$, taking  it  as 
 $H= Q_{-1}/r^{7-p}, \ Q_{-1} \sim  N_{-1} g_s$. 
This prescription may be justified 
 by  the assumption that $N_{-1}$ is large.\foot{By analogy 
with a similar prescription in the type IIA (0-brane)
 case \ci{bbpt}, it may also be given the following heuristic interpretation.
As was noted in \ci{tseprl}, the D-instanton solution of \ci{ggp}
is formally a reduction  of a  gravitational plane wave 
from 12 to 10 dimensions,\ 
\noi
$ds^2_{12} = dx_+ dx_- + K (x) dx_- dx_- + dx_a dx_a$, 
$x_\pm = x_{12} \pm x_{11} $, $a=0,1,..., 9$, $K= Q/r^8$.
 Reducing along
$x_{11}$ and $x_{12}$, i.e.

\noi 
$ds^2_{12} = - e^{-\p} dx^2_{11} + e^\p ( dx_{12} + C_0 dx_{11})^2 
dx_a dx_a $,

\noi
one finds the instanton background
$ e^\p = H=1 + K , \ C_0 = H\inv  -1 $ with the string-frame
meric
$ds^2_{10} = H^{1/2} dx_a dx_a$. 
If instead one reduces along $x_-, x_+$ 
one finds $ e^\p = K , \ C_0 = K\inv , $ \ 
 $ ds^2_{10} = K^{1/2} dx_a dx_a$, i.e. the background with $H\to H-1$
(see also \ci{bps} for a discussion of such shifts in harmonic functions
in connection with T-duality).
Since $H\approx K$ at small $r$ this  background may be interpreted
as a short-distance limit of the D-instanton solution.  Given that 
 the latter
  represents a wormhole
\ci{ggp}, 
this new string-frame metric is flat everywhere, 
 $ ds^2_{10} = Q^{1/2} r^{-4}  dx_a dx_a = Q^{1/2} (d\rho^2 + \rho^2 d \Omega^2_9), \ \rho= 1/r.$}

Since the $F^4$ term in the expansion of the BI action (5) is 
given by the $C_4$ combination, 
 the  resulting leading  term in the interaction potential 
is found to have the same  structure  and  the coefficient
as in   \rf{hhh},\rf{ess}, i.e.  \ci{CT2}
$$ \V^{-1} = - {a_p\td V_{p+1} \ov 8 r^{7-p} } n_{-1} N_{-1} [ \FF^4  - \four (\FF^2)^2]\ , 
$$
where $\FF_{ab} \equiv \F^{-1}_{ab}$.
The fact that the two  abelian 
field  strengths  appearing  in the supergravity
and the  SYM descriptions are related by the 
{\it  inversion}  is  a  consequence 
of  T-duality. T-duality   transforms the $(-1)$ -- $(p +... + (-1))$ system
on $T^p$ into a system of  `pure' 
Dp-brane and Dp-brane with extra charges
$p$ -- $( (-1) + ... +p)$  on  $\td T^p$ which  
is  expected to  have  the $U(n_{-1} + N_{-1})$ SYM theory description.

Closely related expressions are found in the type IIA case
of 0-brane interacting with $p+...+0$ non-marginal bound state. 
The $(p+...+0)$ probe 
action in the 0-brane background  is 
$$
I_p = - T_p  V_p \int dt\  H_0\inv  $$ $$ 
\times\big[ \sqrt { (1 - H_0 v^2)  \det( { H_0^{1/2} \delta_{mn} 
+ \F_{mn} }) }  - \sqrt{ \det {  \F_{mn} }} \big],  $$
where $\F_{mn}$ describes other  brane charges, 
e.g., $n_0 = n_p  (2\pi)^{-p/2} V_p  \sqrt{\det\ \F_{mn}} $.
This action 
may be rewritten as 
$$
I_p = - T_0  
\int dt \  H_0\inv  $$ \be \times \big[ \sqrt { 1 - H_0 v^2} \sqrt{
\det ( { \delta_{mn}  + H_0^{1/2}\FF_{mn} }) }  - 1 \big]  ,  
\la{uuu}
\ee
where $T_0 = n_0 g_s\inv (2\pi)^{1/2}$ and 
$\FF_{mn} \equiv (\F_{mn})\inv$.
In this form it   corresponds to a T-dual  configuration, i.e. 
 to  the interaction of a $p$-brane source (with charge $N_0$) 
with  parallel 
$(0+...+p)$-brane probe  (with  0-brane charge $n_0$) 
 moving in a relative transverse direction.
Introducing  the velocity component 
$\FF_{09}= v$  we can put this action in the same
BI  form as in the above type IIB  example,
$$
 I_p = - T_0 \int dt  H_0\inv  
\big[ \sqrt {-
\det ( { \eta_{ab}  + H_0^{1/2}\FF_{ab} })}  - 1 \big] $$
were again $H_0= Q_0/r^{7-p}$  so that 
 the agreement  between the leading-order long-distance interaction potential
and the SYM result \rf{yyy} is manifest.

Next, one may consider a $p+...+0$  or $p+...+(-1) $ brane probe described
by a Dp-brane action with a constant $\F_{mn}$ field strength 
moving in  type IIA or type IIB 
supergravity backgrounds 
 produced by  a 1/4 or 1/8 supersymmetric  
marginal (or non-marginal) bound state of branes. 
Since the latter are known explicitly (see, e.g., \ci{ts4})
the  computation of the interaction  potentials 
is straightforward.
For example, 
the potentials in the case of 0-brane 
interactions with 1/4 or 1/8 supersymmetric  marginal bound states
have the  universal  form of \rf{pou}.
In the $0-(4\pa0)$ case 

\noi
$ I_0 = -T_0 \int dt   H_0^{-1} \big( \sqrt {1- H_0 H_4 v^2  }
    -   1\big)$, 

\noi
where $H_0 = Q_0/r^3, \ H_4 = 1 + Q_4/r^3, \ Q_4 \sim N_4 g_s $, 
so that 
$$\V^{(1)} =  - { T_0 \ov 8r^{3}}   ( 4 Q_4 v^2 +  
  Q_0  v^4 )
$$
$$ =\   -\frac{n_0}{16r^3} \big( 4 v^2 N_4   +   v^4 N_0\big), $$
which is equivalent   to  the SYM result \rf{ff}  \ci{CT1}. 
Had we kept the constant 1 in  $H_0$ we would get $N_0 + 2 N_4$
as  the coefficient of the 
$v^4$ term   and would need to assume that $N_0 \gg N_4$ to get the 
 agreement with the  SYM  result.

Similarly, using the explicit form of the 
 $4\bot4\bot4\pa 0$ background \ci{KT} one finds that the 
$0-(4\bot4\bot4\pa 0)$ interaction is described by $I_0=$

\noi
$ -T_0 \int dt  H_0^{-1} \big( 
\sqrt {1- H_0 H_{4(1)} H_{4(2)}H_{4(3)} v^2  }  -   1\big)$, 

\noi
where  $H_{4 (k)} = 1 + Q_{4(k)}/r$, so that 
$$
\V^{(1)} = -\frac{n_0}{16r} \big[4 v^2 ( N_{4(1)} + N_{4(2)} +  N_{4(3)})
  +   v^4 N_0\big] , $$
which is again in agreement with the SYM expression \rf{two}.

To summarize, 
the  SYM--supergravity correspondence 
observed   on the above   examples   is formally due to 
   (i)  the   BI-type  structure
of the actions of the non-marginal bound state  branes, \ 
 (ii) 
 the 
`product of harmonic functions' structure 
of the actions in the marginal bound state case
(implying additive dependence of the leading-order potential on constituent charges),
 \ and (iii)  a  combination of these 
two features in 
more general cases of interactions  with  non-marginal
1/4 or 1/8  supersymmetric  bound states.

\section{Subleading term in interaction potentials}
The result of \ci{bbpt}  may be interpreted as  implying that 
the subleading term $\V^{(2)}\sim n_0 N_0^2 g_s v^6/r^{14} $
 in the 0-brane - 0-brane interaction potential \rf{vvv}
in 

\noi
$I_0 =  -T_0 \int dt   H_0^{-1} \big( 
\sqrt {1- H_0  v^2  }  -   1\big)$ 

\noi
is reproduced by the leading 2-loop term  in the  $D=1$ 
SYM effective action $\G$  defined in section 2.1.\foot{Similar SYM interpretation  should  apply to the discussion in 
\ci{GGR}.} This is easy to check by 
assuming that the 2-loop coefficient $\hat C_6$ in $\G$ 
 has the same structure as the 1-loop one $\hat C_4$, i.e.
is given by 
the (symmetrized) trace in the adjoint representation of  the $F^6$ 
term appearing in the expansion of the BI action \rf{see},
\be
\hat C_6 = \STr\ C_6 (F) \ .  \la{ans}
\ee
Indeed, interpreting the velocity as an electric field component 
in $D=2$ SYM theory  and  substituting 
$F_{09} = v J_0$ into \rf{ans}, using that $\Tr J^{2n} = 2 n_0 N_0$
and separating the $n_0 N^2_0$-term  as required \ci{bbpt}  to match 
the  supergravity result (obtained by the probe method)
one finds the  precise agreement with the supergravity potential.

In \ci{CT3} we attempted to test the ansatz \rf{ans} 
by studying the subleading terms in the potentials 
in more complicated  examples of 0-brane interacting with 
bound states of branes wrapped over tori.
Since an explicit 
computation of the  
2-loop term in $\G$ for arbitrary non-abelian gauge field in 
a higher-dimensional SYM  looks as a   complicated problem, we 
followed an indirect route: making a plausible ansatz for the 
2-loop term in $\G$ and then trying  to compare it with  
the supergravity expressions for $\V^{(2)}$ on different 
examples with varied 
amount of supersymmetry, {\it assuming} 
 that  the supergravity-SYM  correspondence should continue
 to hold beyond the leading order as it does in the simplest 
0-brane scattering case. The  consistency
of the resulting picture
 supports   the basic assumption.

The first  non-trivial  example 
is the  0-brane interaction with 1/2 supersymmetric
  non-marginal 
bound state  $(p+...+0)$. Since the  action \rf{uuu}
has the BI structure, the subleading term in its potential part
(the one which is quadratic in  
$H_0 = Q_0/r^{7-p}$) has  the $C_6\sim F^6$ form. 
Plugging  the corresponding  SYM 
 background $F_{09} = v J_0, \ 
F_{mn} = \FF_{mn} J_0$  into \rf{ans} we  find 
the precise agreement  between the SYM and the supergravity expressions
for $\V^{(2)}$  for 
{\it arbitrary} $\F_{mn}$. 

This, in fact,  may be considered  as a  motivation for  choosing \rf{ans}
in the first place. 
A test then comes from the  cases of  0-brane interaction with 
1/4 supersymmetric $1\pa 0$ and $4\pa 0$ bound states.
Though the  supergravity 
action $I_0= -T_0 \int dt   H_0^{-1} \big( \sqrt {1- H_0 H_1 v^2  }
    -   1\big)$
 and thus $\V^{(2)} = - { 1 \ov 16 r^{12} } Q_0 ( 4 Q_1 v^4 + Q_0 v^6)$ 
in the $0-(1\pa 0)$ 
case     have  different structure than 
\rf{uuu}, $\V^{(2)}$ is still reproduced \ci{CT3}
 by the 2-loop term  in $\G$ after one 
 substitutes the 
relevant  $SU(n_0 + N_0)$ background 

\noi
$ 
   F_{09} = v J_0  , \    F_{12} =    F_{02} =h(\td x_1 + x_0) 
 J_0 $

\noi
 into  $\hat C_6$  given by \rf{ans}.

In the $0-(4\pa 0)$  case the supergravity  potential 
has the same form as in the $0-(1\pa 0)$ case, but
the  corresponding SYM background \rf{trat} is now more complicated: it is 
parametrised by two independent commuting matrices $J_0$ and $J_1$. 
Substituting $    F_{09} = v J_0 ,  \ 
   F_{12}=   F_{34} = q J_1$ into \rf{ans}
  gives 

\noi
$\Tr C_6  = - { 1\ov 8}   n_0 N_0 (
 2 v^4 q^2       + v^6 )$ 

\noi
instead of

 $\hat C_6= - { 1\ov 8}   n_0 N_0 (
 4 v^4 q^2       + v^6 )$

\noi
which is needed for agreement with the supergravity potential.

It is natural to  try to modify the ansatz \rf{ans}
in  order to correct the   factor of 2 
discrepancy  in the $v^4$ term,  
without changing, however, 
the result for $\hat C_6$ in all the previous cases (which were 
represented by  more primitive 
 gauge field backgrounds depending on a single 
$SU(n_0 + N_0)$  matrix $J_0$). 
Remarkably, it turns out \ci{CT3} 
that there is a {\it unique }
way of  achieving that goal  (up to  terms  involving commutators of $F$
which are discussed below and 
which vanish on the backgrounds we considered so far):
one is to keep the same Lorentz-index structure of the  $F^6$ terms 
as in $C_6$  but should replace the internal index 
trace $\STr$ by a different combination
of $\tr$, $(\tr)^2$ and $(\tr)^3$ terms (all such
fundamental trace structures 
 may in general appear at 2 loops)\foot{Since here  
we consider only commuting $F_{ab}$
backgrounds,  the symmetrisation  does not play any role.
It is, however, useful  in order to isolate the terms that 
do not vanish in the abelian limit from 
additional commutator terms (see below).}
$$
    \SSTr (Y_{s_1}  ... Y_{s_6} )  \equiv
2N \tr [Y_{(s_1}  ... Y_{s_6)} ] $$ $$ 
 + \ \a_1  \tr [Y_{(s_1}  ... Y_{s_4}] \tr [Y_{s_5} Y_{s_6)}] $$
$$
 +\  \a_2 \tr [Y_{(s_1}  ... Y_{s_3}] \tr [ Y_{s_4}... Y_{s_6)}] $$ $$ 
+\   \a_3 N\inv  \tr [Y_{(s_1}  Y_{s_2}] \tr [ Y_{s_3} Y_{s_4}]
\tr [ Y_{s_5} Y_{s_6)}]. 
$$
$Y_s$ are the $SU(N)$  generators and 
$\a_1= 30, \ \a_2= -20, \ \a_3=0$   in the case when  $\SSTr Y^6 = \STr Y^6$, 
but 
  we need to choose $\a_1= 60, \ \a_2= -50, \ \a_3=-30$ 
in order for  the modified ansatz 
\be
\hat C_6  = \SSTr\ C_6 (F) \ , \la{anz}
\ee
to  reproduce  the supergravity expressions 
in all of the  above examples, 
 including  the $0-(4\pa 0)$ one
(which is the only  case among them 
where $\SSTr C_6(F) \not=\STr C_6(F)$).
Indeed, one finds that for the  gauge field  background 
representing the $0-(4\pa 0)$ configuration 
$$\hat C_6  = \SSTr\ C_6 =- { 1\ov 8(n_0 + N_0) }  
n_0 N_0 $$
$$ \times  \big[ (n_0 + N_0 )  ( 
 2 v^4 q^2       + v^6 ) +   2  N_0  v^4 q^2 \big] , 
$$
so that  the relevant $n_0 N_0^2$ term 
in the 2-loop coefficient $(n_0 + N_0) \hat C_6$ in $\G$
is  now in agreement with the supergravity expression for the 
subleading potential $\V^{(2)}$. 

A  test  of the   consistency of \rf{anz} 
is provided by further 
examples of 0-brane  interactions  with 1/8 supersymmetric 
bound states. In the $0 -(4\bot 1 \pa 0)$ case 
the subleading term in the supergravity potential \rf{pou},\rf{vvv}
is 
$$
\V^{(2)} =   - { T_0 \ov 16 r^{4}}  \big[   8 v^2 Q_4 Q_1 +  4 v^4( Q_4 + Q_1  ) Q_0 
  +   v^6 Q_0^2  \big],  $$
where $Q_1 = g_s Q_0  {N_1\ov N_0}$. 
The same expression  should be  reproduced by the 2-loop SYM effective action
$\G$  with the coefficient 
$\hat C_6$ in \rf{anz}  computed for  the background  \rf{tat}.
One finds that 
$$
\hat C_6  =  - { 1\ov 8(n_0 + N_0) }   n_0 N^2_0 \big[ 16 v^2  w^2 q^2 
$$
$$ + \
 4 v^4  ( q^2+  h^2 +w^2)    + v^6 \big]  + O(n_0^2), 
$$
so that the $v^4$ and $v^6$ terms are indeed the same as 
in $\V^{(2)}$ for any distribution of the total  momentum
between the vector and scalar oscillations 
(representing the momentum wave along the D-string bound to D5-brane in the 
T-dual configuration $5\pa 1+$wave). 
However,  the coefficient of the  $v^2$ term  is {\it non-vanishing} (cf. \ci{dps}) 
  only if $w\not=0$, i.e. only if the scalar 
background is excited.  The $v^2$ term  has the  required coefficient 
provided we  assume that the momentum 
is distributed equally between the scalar and the vector waves,  i.e.
if  $<h^2>=<w^2> = \ha g_s {N_1\ov N_0}$.

Finally, in the $0-(4\bot4\bot4\pa 0)$ case 
the $v^4$ and $v^6$ terms in the  corresponding 
supergravity potential 
$$
\V^{(2)} = -\frac{n_0g_s }{64(2\pi)^{1/2} r^2} $$
$$\times 
\big[\ha v^2 ( N_{4(1)}  N_{4(2)} + 
 N_{4(1)}N_{4(3)} + N_{4(2)}N_{4(3)})
$$
\be
+\  4 v^4 ( N_{4(1)} + N_{4(2)} +  N_{4(3)})N_0
  +   v^6 N_0^2\big] , \la{ffe} \ee
are correctly reproduced by the $\hat C_6$ in \rf{anz} evaluated 
on the background \rf{weew} supplemented by the velocity component
 $F_{09} = v J_0$.  However, 
the coefficient of the $v^2$ term  in the resulting  expression 
for $\G^{(2)}$  
is vanishing 
for both commuting and non-commuting choice of the matrices $\hh_k$ in 
\rf{weew}.  

One should note, however, that  up to this point 
we were  ignoring 
the terms with commutators of $F_{ab}$ which may, of course, 
be present in the 2-loop  effective action, 
but  which were vanishing on the commuting 
backgrounds we were discussing above.  
 In general, one should expect, 
therefore, that\foot{Though this  does not seem 
to be  directly related to the present discussion, 
let us note that, in general, the $F^6$-part  of  the 
 {\it tree-level} (disc) open string theory  effective action 
should also contain certain commutator terms in addition
 to the symmetrised trace terms 
$\Str C_6(F)$  in the non-abelian BI action as defined in 
\ci{nbi}. That such commutator terms may  be needed
in  D-brane  action applications  is  implied by the results of
\ci{hashi,dou,doul}.}
\be
\hat C_6 = \SSTr\ C_6 (F) + \C_6 \ , \ \ \  \C_6= O(F^4[F,F]).
\la{azz}
\ee
To  demonstrate  that the commutator terms $\C_6$ 
can,  indeed,  produce  the needed $v^2$ term,  let us  consider, e.g., 
$$\C_6  \sim 
  \Tr (F_{ab} F_{ab} [F_{cd},F_{ef}] F_{cd} F_{ef} ) . $$
  Making the non-commutative choice of the background in \rf{weew}, 
i.e. taking  $\hh_k$ 
to be  proportional to the Pauli matrices, 
one finds \ci{CT3} that $\C_6$ 
(multiplied by $N=n_0 + N_0$   which is its coefficient in $\G$) 
contains  indeed the  same $v^2$ 
contribution  as in  \rf{ffe}, i.e.  the one 
 proportional to 

\noi
 $ n_0 N_0^2 v^2( N_{4(1)}  N_{4(2)} + 
 N_{4(1)}N_{4(3)} + N_{4(2)}N_{4(3)} )$.

\section{Concluding remarks}
As we have discussed above, 
the supergravity-SYM (matrix theory) 
correspondence  is manifest  for  the leading term in the long-distance interaction potential between 
appropriate configurations 
of branes in $D=10$  (having  large 0-brane number  $N_0$ 
or   finite $N_0$ but 
obtained from $D=11$  using  `null' reduction). 

We have suggested that this  
correspondence holds  also   for  the subleading terms
in the long-distance potential between extended branes, 
  i.e.  not only
for  the $D=1$ SYM (0-brane scattering) case considered in
 \ci{bbpt}. 
It would be  important  to 
perform  a string-theory computation of the subleading  (2-loop) terms 
in the interaction potential, checking  that the $r\to 0$ and $r\to \infty$
limits of the string result continue to agree (for relevant 
configurations of branes) beyond the  leading 1-loop level considered  in \ci{bachas,dkps,lifmat}. This would provide an explanation for
the supergravity-SYM correspondence at the subleading level 
  observed   in \ci{bbpt,CT3}.  

\medskip
\centerline {\ \bf Acknowledgments}
I  am grateful to   K. Becker, M. Becker,  I. Chepelev 
and J. Polchinski  
 for collaborations and useful 
discussions.  I  would like  also to thank
the organizers for the invitation to give a  talk.
This work was supported in part by  
 PPARC and  the European
Commission TMR programme grant ERBFMRX-CT96-0045.

\end{document}